\documentclass[journal]{IEEEtran}
\usepackage{enumerate}
\usepackage{cite}
\usepackage{amsthm}
\usepackage{amsmath}
\usepackage{amscd}
\usepackage{dsfont}
\usepackage{amssymb}
\usepackage{latexsym}
\usepackage{array}
\usepackage{tikz}\usetikzlibrary{calc}
\usepackage{tabularx}
\usepackage{epsfig}
\usepackage{graphicx,subfigure}
\usepackage{epstopdf}
\usepackage[numbers, square, comma, compress]{natbib}
\usepackage{bm}
\newtheorem{lem}{Lemma}
\newtheorem{corol}{Corollary}
\newtheorem{ther}{Theorem}
\newtheorem{deft}{Definition}
\newtheorem{prop}{Proposition}

\newtheorem{exap}{Example}

\theoremstyle{definition}
\newtheorem{rem}{Remark}
\newtheoremstyle{dotless}{}{}{\itshape}{}{\bfseries}{}{ }{}
\theoremstyle{dotless}

\newcommand{\I}{\mathbb{I}}

\newcommand{\V}{\mathbb{V}}

\newcommand{\R}{\mathbb{R}}

\newcommand{\Y}{\mathsf{Y}}
\newcommand{\D}{\mathbb{D}}
\newcommand{\h}{\mathbb{H}}
\newcommand{\p}{\mathbb{P}}
\newcommand {\aplt} {\ {\raise-.5ex\hbox{$\buildrel<\over{\mbox{\scriptsize $\sim$}}$}}\ }
\providecommand{\abs}[1]{\ensuremath{\left\lvert #1 \right\rvert}}
\providecommand{\norm}[1]{\ensuremath{\left\Vert #1 \right\Vert}}

\renewcommand{\IEEEQED}{\IEEEQEDopen}

\DeclareMathOperator*{\Mod}{mod}

\DeclareMathOperator{\SNR}{\mathsf{SNR}}
\newcounter{mytempeqncnt}

\begin{document}

\title{Achieving AWGN Channel Capacity \\With Lattice Gaussian Coding}

\author{Cong~Ling and Jean-Claude~Belfiore
\thanks{%
This is the authors' own version of a paper published in IEEE Trans. Inform. Theory, vol. 60, no. 10, pp. 5918-5929, Oct. 2014.}
\thanks{%
This work was presented in part at the IEEE International Symposium on
Information Theory (ISIT 2013), Istanbul, Turkey, July 2013. This work was supported in part by FP7 project PHYLAWS (EU FP7-ICT 317562).}
\thanks{%
C. Ling is with the Department of Electrical and
Electronic Engineering, Imperial College London, London SW7 2AZ,
United Kingdom (e-mail: cling@ieee.org).
\par
Jean-Claude Belfiore is with the Department of Communications and Electronics, Telecom ParisTech, Paris, France (e-mail: belfiore@telecom-paristech.fr).}
\thanks{Copyright (c) 2014 IEEE. Personal use of this material is permitted.  However, permission to use this material for any other purposes must be obtained from the IEEE by sending a request to pubs-permissions@ieee.org.}
}

\maketitle

\begin{abstract}
We propose a new coding scheme using only one lattice that achieves the $\frac{1}{2}\log(1+\SNR)$ capacity of the additive white Gaussian noise (AWGN) channel with lattice decoding, when the signal-to-noise ratio $\SNR>e-1$. The scheme applies a discrete Gaussian distribution over an AWGN-good lattice, but otherwise does not require a shaping lattice or dither. Thus, it significantly simplifies the default lattice coding scheme of Erez and Zamir which involves a quantization-good lattice as well as an AWGN-good lattice. Using the flatness factor, we show that the error probability of the proposed scheme under minimum mean-square error (MMSE) lattice decoding is almost the same as that of Erez and Zamir, for any rate up to the AWGN channel capacity. We introduce the notion of good constellations, which carry almost the same mutual information as that of continuous Gaussian inputs. We also address the implementation of Gaussian shaping for the proposed lattice Gaussian coding scheme.
\end{abstract}


\section{Introduction}

A practical, structured code achieving the capacity of the power-constrained additive white Gaussian noise (AWGN) channel is the holy grail of communication theory. Lattice codes have been shown to possess this potential. Poltyrev initiated the study of lattice coding without a power constraint, which led to the notion of AWGN-good lattices \cite{Poltyrev94}. Erez and Zamir dealt with the issue of the finite power constraint using nested lattice codes, where a quantization-good lattice serves as the shaping lattice while the AWGN-good lattice serves as the coding lattice \cite{ErezZamir04}. Despite these significant progresses, major obstacles persist from a practical point of view. The scheme of \cite{ErezZamir04} not only requires a dither which complicates the implementation, but also the construction of a quantization-good lattice nested with an AWGN-good lattice is not solved, to the best of our knowledge.

In this paper, we resolve such issues by employing \emph{lattice Gaussian
coding}, when the signal-to-noise ratio $\SNR>e$ \footnote{This threshold is an artifact of the proof technique, which has been reduced in \cite{DBLP:journals/corr/CampelloLB16}. In fact, a new technique is developed in \cite{DBLP:journals/corr/CampelloLB16}, which shows that the equivalent noise $\left(\alpha-1\right)\mathbf{x}+\alpha\mathbf{w}$ in (\ref{eq:equiv-noise}) is \emph{sub-Gaussian}. The sub-Gaussianity not only reduces the SNR condition to $\SNR>e-1$, but also greatly simplifies the proof. More recently, the SNR condition has been completely removed in \cite{CampelloD17}, however at the cost of using dithering. Polar lattices \cite{polarlatticeJ}, which achieve capacity for any SNR, may be seen as an instantiation of \cite{CampelloD17}.}. More precisely, the code book has a discrete Gaussian
distribution over an AWGN-good lattice. So the remaining problem is the construction of AWGN-good lattices, which is nonetheless beyond the scope of this paper (see e.g., \cite{polarlatticeJ,Pietro13ITA,Pietro13,Sommer08,Sadeghi06} for recent progresses which have approached the Poltyrev capacity). Intuitively, since only shaping is lacking in Poltyrev's technique, the probabilistic shaping inherent with lattice Gaussian distribution will enable it to achieve the AWGN channel capacity.

It is well known that the continuous Gaussian distribution is capacity-achieving on the Gaussian channel. Therefore, it is plausible to design Gaussian-like signalling to approach the capacity. This line of work dates back to Shannon's idea in 1948 \cite{Shannon48unpublished}, where nonuniformly spaced pulse-amplitude modulation (PAM) was used to approximate the Gaussian distribution\footnote{It is possible to show that with MMSE scaling at the decoder, Shannon's signalling scheme is approximately good on the Gaussian channel \cite{ShamaiTalk}.}. The capacity of finite constellations with a Gaussian-like distribution was studed in \cite {WuVerdu10}. Discrete Gaussian signalling over lattices was used in \cite{Forney_Wei_89,Kschischang_Pasupathy,BK:Zamir,Tutorial:Zamir,Palgy12} for shaping over the AWGN channel, and more recently in \cite{LLBS_12} to achieve semantic security over the Gaussian wiretap channel. Our novel contribution in this paper is to use
the flatness factor \cite{LLBS_12} to prove that discrete Gaussian signaling over
AWGN-good lattices can achieve the capacity of the power-constrained Gaussian
channel with minimum mean-square error (MMSE) lattice decoding. The concept of flatness factor relates to the properties of Gaussian measures on lattices, and was first introduced in \cite{BelfioreITW11} in the context of physical-layer network coding. In \cite{LLBS_12}, the authors also showed the relevance of the flatness factor for secrecy coding and introduced the notion of \emph{secrecy-good lattices} for the Gaussian wiretap channel. We note that in \cite{Tutorial:Zamir}, achieving the AWGN channel capacity using non-uniform signaling is posed as an open question. This paper serves as answer to \cite{Tutorial:Zamir} in the affirmative. Furthermore, with the flatness factor, we are able to provide considerable new insights into some existing intuitions and make them rigorous, which were only established in literature under certain approximations. For example, although it is believed that the ultimate shaping gain ($\pi e/6$ or 1.53 dB) can be achieved by the lattice Gaussian distribution for any dimension, it was only derived with the continuous approximation \cite{Forney_Wei_89,Kschischang_Pasupathy}. In this paper, a precise bound on the shaping gain is derived, which converges to 1.53 dB as the flatness factor tends to zero.

The proposed approach enjoys a couple of salient features. Firstly, throughout the paper, we do not use a shaping lattice. Secondly, in contrast to what is nowadays the common practice of lattice coding \cite{ErezZamir04}, we do not use a dither. These will simplify the implementation of the system. In the meantime, compared to Voronoi shaping, the downside of probabilistic shaping is the variable rate of input data, since the constellation points are not equally probable. This side effect warrants further investigation and may be handled by data buffering \cite{Forney_Wei_89,Kschischang_Pasupathy}.

As we will see, the lattice Gaussian distribution behaves like the continuous Gaussian distribution in many aspects, while still preserving the rich structures of a lattice. Since the continuous Gaussian distribution is capacity-achieving for many problems in information theory, we expect lattice Gaussian coding will find more applications, especially in network information theory, where structures of the code are desired for the purpose of coordination.

This paper is organized as follows. In Section II, we review lattice Gaussian distributions and derive new properties of the flatness factor, including the mutual information carried by a lattice Gaussian constellation. This leads to the notion of good constellations in the sense of capacity.
Section III gives the coding theorem for lattice Gaussian coding under MMSE lattice decoding.
Section IV addresses the implementation of lattice Gaussian coding. In Section VI, we conclude the paper with a brief discussion.

Throughout this paper, we use the natural logarithm, denoted by $\log$, and information is measured in nats.

\section{Lattice Gaussian Distribution and Flatness Factor}

In this section, we introduce the mathematical tools needed to describe
and analyze the proposed coding scheme.

\subsection{Preliminaries of Lattice Coding}

An $n$-dimensional {lattice} $\Lambda$ in the Euclidean space
$\mathbb{R}^{n}$ is a set defined by
\begin{equation*}
\Lambda=\mathcal{L}\left( \mathbf{B}\right) =\left\{ \mathbf{Bx}\text{ : }\mathbf{x\in }\text{ }%
\mathbb{Z}^{n}\right\}
\end{equation*}%
where the columns of the basis matrix $\mathbf{B=}\left[
\mathbf{b}_{1}\cdots \mathbf{b}_{n}\right] $ are linearly
independent. (In this work, we will restrict ourselves to full-rank lattices.)

For a vector $\mathbf{x}\in\mathbb{R}^{n}$, the nearest-neighbor quantizer associated
with $\Lambda$ is
$Q_{\Lambda}(\mathbf{x})=\arg\min_{{\bm \lambda} \in\Lambda}\|{\bm \lambda}-\mathbf{x}\|$.
We define the modulo lattice operation by $\mathbf{x} \mod \Lambda \triangleq
\mathbf{x} -Q_{\Lambda}(\mathbf{x})$. The Voronoi cell of $\Lambda$,
defined by
$\mathcal{V}(\Lambda)=\{\mathbf{x}:Q_{\Lambda}(\mathbf{x})=\mathbf{0}\}$,
specifies the nearest-neighbor decoding region. The Voronoi cell is one example of fundamental region of the lattice. A measurable set~$\mathcal{R}(\Lambda)\subset \mathbb{R}^n$ is a fundamental region of the lattice~$\Lambda$ if~$\cup_{\bm{{\bm \lambda}} \in \Lambda} (\mathcal{R}(\Lambda)+\bm{{\bm \lambda}}) = \R^n$ and
if~$(\mathcal{R}(\Lambda)+\bm{{\bm \lambda}}) \cap (\mathcal{R}(\Lambda)+\bm{{\bm \lambda}}')$
has measure~$0$ for any~$\bm{{\bm \lambda}} \neq \bm{{\bm \lambda}}'$
in~$\Lambda$. The volume of a fundamental region is equal to that of the Voronoi cell~$V(\Lambda)=\sqrt{|\det(\mathbf{B}^T \mathbf{B})|}$.

The {theta series} of $\Lambda$ (see, e.g., \cite{BK:Conway98}) is defined as
\begin{eqnarray}
\Theta_{\Lambda}(q)=\sum_{{\bm \lambda} \in \Lambda} q^{\|{\bm \lambda}\|^2}
\end{eqnarray}
where $q= e^{j\pi z}$ ($j=\sqrt{-1}$ and the imaginary part $\Im(z)>0$). Letting $z$ be purely imaginary, and assuming $\tau=\Im(z)>0$, we can
alternatively express the theta series as
\begin{eqnarray}
\Theta_{\Lambda}(\tau)=\sum_{{\bm \lambda} \in \Lambda} e^{-\pi
\tau\|{\bm \lambda}\|^2}.
\end{eqnarray}

Consider the problem of infinite lattice coding over the AWGN channel \cite{Poltyrev94}. Let $\sigma^2$ be the power of the i.i.d.\ Gaussian noise $\mathsf{W}^n$.
For an $n$-dimensional lattice $\Lambda$, define the
volume-to-noise ratio (VNR) \footnote{The definition of VNR varies slightly in literature, by a factor $2\pi$ or $2\pi e$. In particular, the VNR is defined as $V(\Lambda)^{\frac{2}{n}}/(2\pi e\sigma^2)$ in \cite{Forney00,ErezZamir04}.} by
\[
\gamma_{\Lambda}(\sigma)\triangleq\frac{(V(\Lambda))^{\frac{2}{n}}}{\sigma^2}
\]
The error probability of minimum-distance lattice decoding is given by $P_e = \mathbb{P}\{ \mathsf{W}^n \notin
\mathcal{V}(\Lambda)\}$.

Let us introduce the notion of lattices which are good for the
Gaussian channel without a power constraint \cite{BK:Zamir}:

\begin{deft}[AWGN-good lattices]
A sequence of lattices~$\Lambda^{(n)}$ of increasing dimension $n$ is \emph{AWGN-good} if, for all $P_e \in (0,1)$,
\[
\lim_{n \to \infty} \gamma_{\Lambda^{(n)}}(\sigma)=2\pi e
\]
and if, for a fixed VNR greater than $ 2\pi e$, $P_e$ vanishes in $n$.
\end{deft}

Erez and Zamir~\cite{ErezZamir04} showed that lattice coding and decoding can achieve
the capacity of the Gaussian channel. More
precisely, one can prove the existence of a sequence of nested
lattices $\Lambda_s^{(n)} \subset \Lambda_f^{(n)}$ such that
\begin{enumerate}
\item[-] the shaping lattice $\Lambda_s^{(n)}$ is simultaneously quantization-good and AWGN-good;
\item[-] the fine lattice $\Lambda_f^{(n)}$ is AWGN-good.
\end{enumerate}
When a random dither at the transmitter and an MMSE filter at the
receiver are used, the Voronoi signal constellation $\Lambda_f^{(n)}
\cap \mathcal{V}(\Lambda_s^{(n)})$ approaches the
capacity of the Gaussian channel, when $n$ is large (see~\cite{ErezZamir04}).

\subsection{Lattice Gaussian Distribution} \label{lattice_gaussian}

\begin{figure}[t]

\centering\centerline{\epsfig{figure=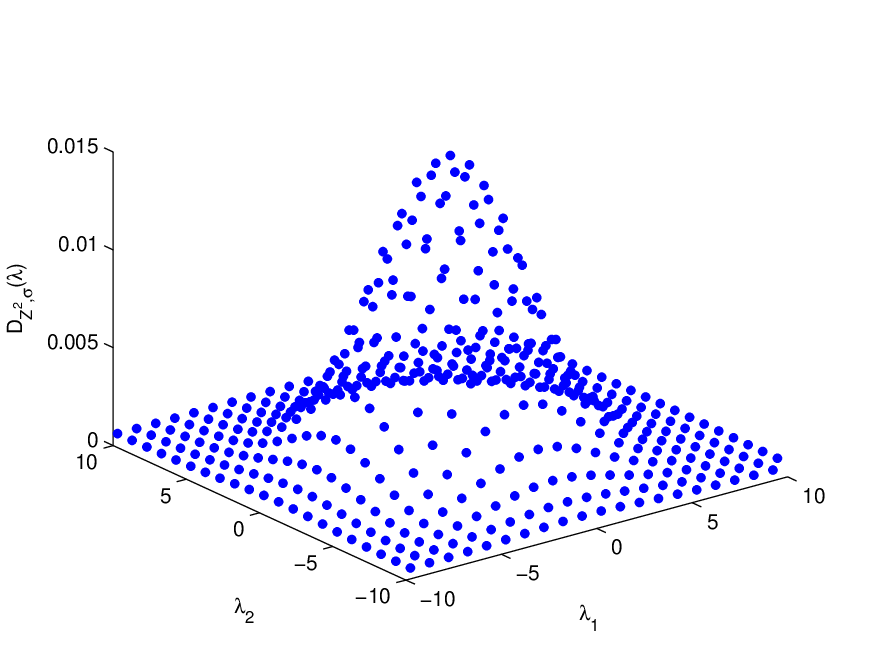,width=9cm}}

\caption{Discrete Gaussian distribution over $\mathbb{Z}^2$. The height represents the probability of a lattice point $D_{\mathbb{Z}^2,\sigma}({\bm \lambda})$ where ${\bm \lambda} = (\lambda_1,\lambda_2)^T \in \mathbb{Z}^2$.}

\vspace{-0.5cm}

\label{fig:Discrete_Gaussian}
\end{figure}

For~$\sigma>0$ and $\mathbf{c} \in \R^n$,
we define the Gaussian distribution of variance $\sigma^2$ centered at ${\bf c} \in \R^n$ as
\begin{equation*}
 f_{\sigma,{\bf c}}(\mathbf{x})=\frac{1}{(\sqrt{2\pi}\sigma)^n}e^{- \frac{\|\mathbf{x}-{\bf c}\|^2}{2\sigma^2}},
\end{equation*}
for all $\mathbf{x} \in\R^n$. For convenience, we write $f_{\sigma}(\mathbf{x})=f_{\sigma,{\bf
0}}(\mathbf{x})$.

We also consider the $\Lambda$-periodic function
\begin{equation}\label{Guass-function-lattice}
  f_{\sigma,\Lambda}(\mathbf{x})=\sum_{{{\bm \lambda}}\in \Lambda}
{f_{\sigma,{{\bm \lambda}}}(\mathbf{x})}=\frac{1}{(\sqrt{2\pi}\sigma)^n}
\sum_{{\bm \lambda} \in \Lambda} e^{-
    \frac{\|\mathbf{x}-{\bm \lambda}\|^2}{2\sigma^2}},
\end{equation}
for all $\mathbf{x} \in\R^n$. Observe that $f_{\sigma,\Lambda}$ restricted to the fundamental region $\mathcal{R}(\Lambda)$ is a probability density.

We define the \emph{discrete Gaussian distribution} over $\Lambda$
centered at $\mathbf{c} \in \R^n$ as the following discrete
distribution taking values in ${\bm \lambda} \in \Lambda$:
\[
D_{\Lambda,\sigma,\mathbf{c}}({\bm \lambda})=\frac{f_{\sigma,\mathbf{c}}(\mathbf{{\bm \lambda}})}{f_{\sigma,\mathbf{c}}(\Lambda)}, \quad \forall {\bm \lambda} \in \Lambda,
\]
where $f_{\sigma,\mathbf{c}}(\Lambda) \triangleq \sum_{{\bm \lambda} \in
\Lambda} f_{\sigma,\mathbf{c}}(\mathbf{{\bm \lambda}})=f_{\sigma,\Lambda}(\mathbf{c})$. Again for convenience, we write $D_{\Lambda,\sigma}=D_{\Lambda,\sigma,\mathbf{0}}$. We remark that
this definition differs slightly from the one in~\cite{Micciancio05}, where~$\sigma$ is scaled by a constant factor $\sqrt{2\pi}$ (i.e., $s=\sqrt{2\pi}\sigma$). Fig. \ref{fig:Discrete_Gaussian} illustrates the discrete Gaussian distribution over $\mathbb{Z}^2$. As can be seen, it resembles a continuous Gaussian distribution, but is only defined over a lattice. In fact, discrete and continuous Gaussian distributions share similar properties, if the flatness factor is small.

It will be useful to define the {discrete Gaussian distribution} over a coset of $\Lambda$, i.e., the shifted lattice $\Lambda-\mathbf{c}$:
\[
D_{\Lambda-\mathbf{c},\sigma}({\bm \lambda}-\mathbf{c})=\frac{f_{\sigma}(\mathbf{{\bm \lambda}}-\mathbf{c})}{f_{\sigma, {\bf c}}(\Lambda)}, \quad \forall {\bm \lambda} \in \Lambda.
\]
Note the relation $D_{\Lambda-\mathbf{c},\sigma}({\bm \lambda}-\mathbf{c}) = D_{\Lambda,\sigma,\mathbf{c}}({\bm \lambda})$, namely, they are a shifted version of each other.

The following lemma due to Banaszczyk \cite{Banaszczyk} shows that each component of ${\bm \lambda}\sim D_{\Lambda,\sigma}$ (i.e., ${\bm \lambda}$ is sampled from distribution $D_{\Lambda,\sigma}$) has an average power always less than $\sigma^2$.

\begin{lem}\label{lem:Banaszczyk}
Let ${\bm \lambda}=(\lambda_1, \lambda_2, \ldots, \lambda_n)^T \sim D_{\Lambda,\sigma}$. Then for each $1\leq k \leq n$
\begin{equation}
\mathbb{E}[{\lambda}_k^2] \leq \sigma^2.
\end{equation}
\end{lem}


%

\subsection{Flatness Factor}

The flatness factor of a lattice~$\Lambda$ quantifies the maximum variation of~$f_{\sigma,\Lambda}(\mathbf{x})$ for~$\mathbf{x} \in \R^n$.

\begin{deft} [Flatness factor \cite{LLBS_12}]
For a lattice~$\Lambda$ and for a parameter~$\sigma$, the flatness factor
is defined by:
\begin{equation*}
\epsilon_{\Lambda}(\sigma)  \triangleq \max_{\mathbf{x} \in
\mathcal{R}(\Lambda)}\abs{
V(\Lambda)f_{\sigma,\Lambda}(\mathbf{x})-1}.
\end{equation*}
\end{deft}

In other words, $\frac{f_{\sigma,\Lambda}(\mathbf{x})}{1/V(\Lambda)}$, the ratio between $f_{\sigma,\Lambda}(\mathbf{x})$ and the uniform distribution over~$\mathcal{R}(\Lambda)$, is within the range $[1-\epsilon_{\Lambda}(\sigma), 1+\epsilon_{\Lambda}(\sigma)]$.

\begin{prop} [Expression of $\epsilon_{\Lambda}(\sigma)$ \cite{LLBS_12}] \label{expression_epsilon}
We have:
\begin{equation*}
\epsilon_{\Lambda}(\sigma) =  \left(\frac{\gamma_{\Lambda}(\sigma)}{{2\pi}}\right)^{\frac{n}{2}}{
\Theta_{\Lambda}\left({\frac{1}{2\pi\sigma^2}}\right)}-1
\end{equation*}
where $\gamma_{\Lambda}(\sigma) = \frac{
V(\Lambda)^{\frac{2}{n}}}{\sigma^2}$ is the VNR.
\end{prop}

Consider the ensemble of mod-$p$ lattices (Construction~A) \cite{Loeliger}. Denote by $\mathbb{Z}_p$ the ring of integers modulo-$p$.
For integer $p>0$, let $\mathbb{Z}^n \rightarrow \mathbb{Z}^n_p:
{\mathbf{v}} \mapsto \overline{\mathbf{v}}$ be the element-wise
reduction modulo-$p$. The mod-$p$
lattices are defined as $\Lambda_C \triangleq
\{\mathbf{v}\in \mathbb{Z}^n: \overline{\mathbf{v}} \in C\}$, where
$p$ is a prime and $C$ is a linear code over $\mathbb{Z}_p$. Quite often, scaled mod-$p$ lattices $a \Lambda_C \triangleq \{a
\mathbf{v}: \mathbf{v}\in \Lambda_C\}$ for some $a \in
\mathbb{R}^+$ are used. The fundamental volume of such a lattice is
$V(a \Lambda_C)=a^n p^{n-k}$, where $n$ and $k$ are the
block length and dimension of the code $C$, respectively.

The following result guarantees the existence of sequences of mod-$p$ lattices whose
flatness factors can vanish
as $n \to \infty$.

\begin{ther}[\cite{LLBS_12}]\label{theorem2}
$\forall \sigma>0$ and $\forall\delta>0$, there exists a sequence of mod-$p$ lattices~$\Lambda^{(n)}$ such that
\begin{equation} \label{eq:flatness_factor_bound}
\epsilon_{\Lambda^{(n)}}(\sigma) \leq
(1+\delta) \cdot \left(\frac{\gamma_{\Lambda^{(n)}}(\sigma)}{2\pi}\right)^{\frac{n}{2}},
\end{equation}
i.e., the flatness factor can go to zero exponentially for any fixed
VNR $\gamma_{\Lambda^{(n)}}(\sigma)<2\pi$.
\end{ther}

%
%


\subsection{Properties of the Flatness Factor}
The importance of a small flatness factor is two-fold. Firstly, it assures the ``folded" distribution $f_{\sigma,\Lambda}(\mathbf{x})$ is flat; secondly, it implies the discrete Gaussian distribution $D_{\Lambda,\sigma,\mathbf{c}}$ is ``smooth".
In this subsection we collect known properties and further derive new
properties of lattice Gaussian distributions that will be useful in
the paper.

From the definition of the flatness factor, one can derive
the following result:

\begin{lem} \label{lem:2}
For all~${\bf c} \in \R^n$ and~$\sigma>0$, we have:
\[
{f_{\sigma,{\bf c}}(\Lambda)} \in {[1- \epsilon_{\Lambda}(\sigma), 1+ \epsilon_{\Lambda}(\sigma)]\frac{1}{V(\Lambda)}}.
\]
\end{lem}




\begin{lem}\label{lemma5}
Let $\Lambda^N$ be the $N$-fold Cartesian product of the lattice $\Lambda$. Then
\[
\epsilon_{\Lambda^N}(\sigma) = [1+\epsilon_{\Lambda}(\sigma)]^N-1.
\]
In particular, $\epsilon_{\Lambda^N}(\sigma) \approx N\epsilon_{\Lambda}(\sigma)$ if $\epsilon_{\Lambda}(\sigma)$ is small.
\end{lem}

\begin{proof}
Use the facts $\Theta_{\Lambda^N}(x)=\Theta^N_{\Lambda}(x)$ and $V(\Lambda^N)=V^N(\Lambda)$ in the definition of the flatness factor.
\end{proof}

\begin{lem}[\cite{LLBS_12}]\label{lemma6}
Let $\Lambda' \subset \Lambda$ be a pair of nested lattices such that~$\epsilon_{\Lambda'}(\sigma)<\frac{1}{2}$. Then
\[
\V(D_{\Lambda,\sigma,\mathbf{c}}\Mod \Lambda',U(\Lambda/\Lambda')) \leq 4\epsilon_{\Lambda'}(\sigma),
\]
where~$U(\Lambda/\Lambda')$ denotes the uniform distribution over the finite set~$\Lambda/\Lambda'$.
Conversely, if $\mathbf{a}$ is uniformly distributed in $\Lambda/\Lambda'$ and $\mathbf{b}$ is sampled from~$D_{\Lambda',\sigma,\mathbf{c-a}}$, then the distribution $D_{\mathbf{a}+\mathbf{b}}$ of $\mathbf{a}+\mathbf{b}$ satisfies
\[
\V(D_{\mathbf{a}+\mathbf{b}},D_{\Lambda,\sigma,\mathbf{c}}) \leq \frac{2\epsilon_{\Lambda'}(\sigma)}{1-\epsilon_{\Lambda'}(\sigma)}.
\]
\end{lem}

The following result shows that the variance per dimension of the
discrete Gaussian $D_{\Lambda,\sigma,\mathbf{c}}$ is not far
from $\sigma^2$ when the flatness factor is small. The proof can be found in
\cite[Appendix III-C]{LLBS_12}.
\begin{lem} [Variance of lattice Gaussian \cite{LLBS_12}] \label{Micciancio_Regev_Lemma43}
Let $\mathbf{x} \sim D_{\Lambda,\sigma,\mathbf{c}}$. If
$\varepsilon = \epsilon_{\Lambda}\left(\sigma/\sqrt{\frac{\pi}{\pi-t}}\right) < 1$ for $0<t<\pi$, then
\[
\abs{\mathbb{E}\left[\norm{\mathbf{x}-\mathbf{c}}^2\right]-n\sigma^2} \leq
\frac{2\pi\varepsilon_t}{1-\varepsilon}\sigma^2
\]
where
\[
\varepsilon_t \triangleq
\left\{
  \begin{array}{ll}
    \varepsilon, & \hbox{$t \geq 1/e$;} \\
    (t^{-4}+1)\varepsilon, & \hbox{$0< t < 1/e$.}
  \end{array}
\right.
\]
\end{lem}

\begin{rem}
Note that the extra coefficient $\sqrt{\frac{\pi}{\pi-t}}\approx 1.06$ when $t=1/e$. It can be further reduced arbitrarily close to 1, at the cost of another constant $t^{-4}+1$ before the flatness factor. Nonetheless, this constant can be compensated by increasing $n$ to make the flatness factor decrease exponentially. So essentially one only needs small  $\epsilon_{\Lambda}(\sigma)$ such that the variance of lattice Gaussian is approximately $\sigma^2$. The condition of negligible $\epsilon_{\Lambda}(\sigma)$ can hold for any $n$ and for any $\Lambda$ as long as $\sigma$ is sufficiently large.
For example, $\epsilon_{\mathbb{Z}}(\sigma) = 3\times 10^{-5}$ when $\sigma=0.75$. Basically, the requirement is that $\sigma$ is larger than the smoothing parameter \cite{Re09}.
\end{rem}

From the maximum-entropy principle \cite[Chap.~11]{BK:Cover}, it
follows that the discrete Gaussian distribution maximizes the
entropy given the average energy and given the same support over a lattice.
This is still so even if we
restrict the constellation to a finite region of a lattice.
The following lemma further shows that if the flatness factor is
small, the entropy rate of a discrete Gaussian
$D_{\Lambda,\sigma,\mathbf{c}}$ is almost equal to the differential
entropy of a continuous Gaussian of variance~$\sigma^2$, minus $\frac{1}{n}\log V(\Lambda)$,
that of a uniform distribution over the fundamental region of $\Lambda$.

\begin{lem}[Entropy of lattice Gaussian \cite{LLBS_12}] \label{proposition4}
Let $\mathbf{x} \sim
D_{\Lambda,\sigma,\mathbf{c}}$. If $\varepsilon = \epsilon_{\Lambda}\left(\sigma/\sqrt{\frac{\pi}{\pi-t}}\right) < 1$ for $0<t<\pi$,
then the entropy rate of~$\mathbf{x}$ satisfies
\[
\abs{\frac{1}{n}\mathbb{H}(\mathbf{x}) - \left[\log (\sqrt{2 \pi e}\sigma) - \frac{1}{n}\log {V(\Lambda)}\right]} \leq \varepsilon',
\]
where $\varepsilon'= - \frac{\log(1-\varepsilon)}{n} + \frac{\pi\varepsilon_t}{n(1-\varepsilon)}$.
\end{lem}

Combining Lemmas \ref{Micciancio_Regev_Lemma43} and \ref{proposition4}, we can show that the lattice Gaussian distribution enjoys the optimum shaping gain (1.53 dB) when the flatness factor is small. Note that our proof does not require the continuous approximation in \cite{Forney_Wei_89,Kschischang_Pasupathy}, where a discrete Gaussian distribution was intuitively approximated by a continuous one. The following new lemma makes this intuition precise.

\begin{lem}[Shaping gain of lattice Gaussian] \label{lemma:shaping-gain}
Consider lattice Gaussian distribution $D_{\Lambda-\mathbf{c},\sigma}$ for any $\mathbf{c} \in \mathbb{R}^n$. If $\varepsilon = \epsilon_{\Lambda}\left(\sigma/\sqrt{\frac{\pi}{\pi-t}}\right) < 1$ for $0<t<\pi$,
then it shaping gain is bounded by
\[
\gamma_s \geq \frac{\pi e}{6} \cdot \frac{2^{-2\varepsilon'}}{1 + \frac{2\pi\varepsilon_t}{n(1-\varepsilon)}}
\]
where $\varepsilon'= - \frac{\log(1-\varepsilon)}{n} + \frac{\pi\varepsilon_t}{n(1-\varepsilon)}$.
In particular, $\gamma_s \approx \frac{\pi e}{6}$ ($1.53$ dB) if $\varepsilon$ is negligible.
\end{lem}

\begin{proof}
By Lemma \ref{Micciancio_Regev_Lemma43}, if $\mathbf{x} \sim D_{\Lambda-\mathbf{c},\sigma}$, then its power per dimension is upper-bounded by
\begin{equation}\label{eq:power-upper-bound}
\frac{1}{n}\mathbb{E}\left[\norm{\mathbf{x}}^2\right] \leq \sigma^2 + \frac{2\pi\varepsilon_t}{n(1-\varepsilon)}\sigma^2.
\end{equation}
By Lemma \ref{proposition4}, its entropy rate is lower-bounded by
\begin{equation}
\frac{1}{n}\mathbb{H}(\mathbf{x}) \geq \log (\sqrt{2 \pi e}\sigma) - \frac{1}{n}\log {V(\Lambda)} - \varepsilon'.
\nonumber
\end{equation}
Following the footsteps of \cite{Forney_Wei_89}, we know the baseline power (the power for a cubic shaping region over the same coding lattice $\Lambda$) per dimension for this bit rate is
\begin{equation}\label{eq:power-lower-bound}
\frac{\left(2^{\frac{1}{n}\mathbb{H}(\mathbf{x})}V(\Lambda)^{1/n}\right)^2}{12} \geq \frac{2 \pi e\sigma^2 \cdot 2^{-2\varepsilon'}}{12}.
\end{equation}
The shaping gain is defined as the ratio between the baseline power and the actual power:
\[
\gamma_s = \frac{\frac{\left(2^{\frac{1}{n}\mathbb{H}(\mathbf{x})}V(\Lambda)^{1/n}\right)^2}{12}}{\frac{1}{n}\mathbb{E}\left[\norm{\mathbf{x}}^2\right]. }
\]
Using (\ref{eq:power-lower-bound}) and (\ref{eq:power-upper-bound}), we obtain the lower bound on $\gamma_s$ in the lemma.
\end{proof}

The next lemma shows that a sample from a discrete Gaussian distribution with parameter
$\sigma$ is at most $\sqrt{n}\sigma$ away from its center with high probability.
The proof is given in Appendix~\ref{appendix0}.

\begin{lem}[]\label{lem:sphere}
Let $\mathbf{x} \sim
D_{\Lambda,\sigma,\mathbf{c}}$ and
$\varepsilon = \epsilon_{\Lambda}(\sigma)<1$. Then for any $\rho> 1$, the probability
\begin{equation}
\p(\|\mathbf{x}-\mathbf{c}\|>\rho \cdot \sqrt{n}\sigma) \leq \frac{1+\varepsilon}{1-\varepsilon} \cdot e^{-nE_{\mathrm{sp}}(\rho^2)}
\end{equation}
where $E_{\mathrm{sp}}(x)=\frac{1}{2}[x-1-\log(x)]$ for $x>1$ is the sphere-packing exponent.
\end{lem}

This lemma extends \cite[Lemma 4.4]{Micciancio05}, which states that $\p(\|\mathbf{x}-\mathbf{c}\|> \sqrt{2\pi n}\sigma)<\frac{1+\varepsilon}{1-\varepsilon} \cdot 2^{-n}$.

It is well known that the probability of the continuous Gaussian distribution falling outside of a ball of radius larger than $\sqrt{n}\sigma$ is exponentially small. Interestingly, Lemma \ref{lem:sphere} shows this property also holds for the lattice Gaussian distribution, with the same sphere-packing exponent \cite{BK:Zamir}.

Following the definition of the generalized asymptotical equipartition property (AEP) in \cite{BK:Zamir}, we generalize the AEP of i.i.d. Gaussian vectors to the lattice Gaussian distribution.

\begin{prop}[Generalized AEP]\label{prop:typicality}
Let $\mathbf{x} \sim
D_{\Lambda,\sigma,\mathbf{c}}$. If
$\epsilon_{\Lambda}(\sigma) \to 0$, then $\mathbf{x}$ satisfies the generalized AEP, namely,
%
\begin{enumerate}
  \item
\[
\frac{1}{n} \h(\mathbf{x}) \to \frac{1}{2}\log(2\pi e) - \frac{1}{2}\log(\gamma_{\Lambda}(\sigma));
\]
  \item For any $\widetilde{\varepsilon}>0$, there exists a typical set  $T_{\widetilde{\varepsilon}}^{(n)} =\left\{ \mathbf{x}\in \Lambda : \| \mathbf{x}-\mathbf{c} \| \leq \rho(\widetilde{\varepsilon})\sqrt{n}\sigma \right\}$ where $\rho(\widetilde{\varepsilon}) \gtrapprox 1$ such that
   \[
      \p(\mathbf{x} \in T_{\widetilde{\varepsilon}}^{(n)}) > 1-\widetilde{\varepsilon};
      \]
  \item The size of the typical set is approximately $(2\pi e \sigma^2)^{n/2} /V(\Lambda)$.
\end{enumerate}
\end{prop}

The proof is straightforward: Item 1) follows from Lemma \ref{proposition4}; Item 2) is due to Lemma \ref{lem:sphere} and the fact that $\rho(\widetilde{\varepsilon}) \to 1$ as $n \to \infty$; Item 3) is the number of lattice points in a ball of radius $\sqrt{n} \sigma$.

The following lemma by Regev (adapted from~\cite[Claim~3.9]{Re09})
shows that if the flatness factor is small, the sum of a discrete
Gaussian and a continuous Gaussian is very close to a continuous
Gaussian.
\begin{lem} \label{Regev_Claim39}
Given any vector $\mathbf{c} \in \R^n$, and $\sigma_s,\sigma>0$. Let $\tilde{\sigma}\triangleq \frac{\sigma_s\sigma}{\sqrt{\sigma_s^2+\sigma^2}}$ and let $\sigma_s'=\sqrt{\sigma_s^2+\sigma^2}$. Consider the continuous
distribution~$g$
on~$\R^n$ obtained by adding a continuous Gaussian
of variance $\sigma^2$ to a discrete
Gaussian~$D_{\Lambda-\mathbf{c},\sigma_s}$:
\[
g(\mathbf{x})=\frac{1}{f_{\sigma_s}(\Lambda-\mathbf{c}) } \sum_{\mathbf{t} \in \Lambda -\mathbf{c}} f_{\sigma_s}(\mathbf{t}) f_{\sigma}(\mathbf{x}-\mathbf{t}), \quad \mathbf{x} \in \R^n.
\]
If~$\varepsilon =
\epsilon_{\Lambda}\left(\tilde{\sigma}\right)
<\frac{1}{2}$,
then $\frac{g(\mathbf{x})}{f_{\sigma_s'}(\mathbf{x})}$ is uniformly close to $1$:
\begin{equation} \label{Regev_uniform_convergence}
\forall \mathbf{x} \in \R^n, \quad
\abs{\frac{g(\mathbf{x})}{f_{\sigma_s'}(\mathbf{x})}-1}
\leq 4 \varepsilon.
\end{equation}
\end{lem}

\begin{rem}
Interestingly, if $\sigma_s^2$ and $\sigma^2$ respectively represent the signal and noise variances, then $\tilde{\sigma}^2$ can be interpreted as the noise variance scaled by the MMSE coefficient, since by Lemma \ref{Micciancio_Regev_Lemma43}, $\sigma_s^2$ is the signal power as the flatness factor tends to zero.
\end{rem}

\begin{corol}\label{corol-varational}
If~$\varepsilon =
\epsilon_{\Lambda}\left(\tilde{\sigma}\right)
<\frac{1}{2}$, the variational distance between $g(\mathbf{x})$ and the continuous Gaussian density $f_{\sigma_s'}$ is bounded as
\[
\V\left(g,f_{\sigma_s'}\right)\leq 4\varepsilon.
\]
\end{corol}

\begin{corol}\label{corol-KL}
If~$\varepsilon =
\epsilon_{\Lambda}\left(\tilde{\sigma}\right)
<\frac{1}{2}$, the Kullback-Leibler divergence between $g(\mathbf{x})$ and the continuous Gaussian density $f_{\sigma_s'}$ is bounded as
\[
\D\left(g,f_{\sigma_s'}\right)\leq \log(1+4\varepsilon).
\]
\end{corol}

\begin{proof}
\begin{align*}
\D\left(g,f_{\sigma_s'}\right) &= \int_{\R^n} g(\mathbf{x}) \log
\frac{g(\mathbf{x})}{f_{\sigma_s'}(\mathbf{x})} d\mathbf{x}\\
&\stackrel{(a)}{\leq} \int_{\R^n} g(\mathbf{x}) \log(1+4\varepsilon) d\mathbf{x} \\
&= \log(1+ 4\varepsilon)
\end{align*}
where (a) is due to Regev's uniform convergence~(\ref{Regev_uniform_convergence}).
\end{proof}

Regev's lemma leads to an important property, namely, the discrete
Gaussian distribution over a lattice preserves the capacity of the AWGN channel if
the flatness factor is negligible. The proof of the following theorem is given in Appendix \ref{appendixI}.

\begin{ther}[Mutual information of discrete Gaussian distribution]\label{theorem:capacity}
Consider an AWGN channel where the input constellation $\mathsf{X}$ has
a discrete Gaussian distribution $D_{\Lambda-\mathbf{c},\sigma_s}$
for arbitrary $\mathbf{c} \in \mathbb{R}^n$, and where the variance
of the noise $\mathsf{W}$ is $\sigma_w^2$. Let the average signal power be $P$ so that $\SNR=P/\sigma_w^2$, and let $\tilde{\sigma}_w\triangleq \frac{\sigma_s\sigma_w}{\sqrt{\sigma_s^2+\sigma_w^2}}$. Then, if
$\varepsilon = \epsilon_{\Lambda}\left(\tilde{\sigma}_w\right) < \frac{1}{2}$ and $\frac{\pi\varepsilon_t}{1-\epsilon_t}\leq \varepsilon$ where
\[
\varepsilon_t \triangleq
\left\{
  \begin{array}{ll}
    \epsilon_{\Lambda}\left(\sigma_s/\sqrt{\frac{\pi}{\pi-t}}\right), & \hbox{$t \geq 1/e$} \\
    (t^{-4}+1)\epsilon_{\Lambda}\left(\sigma_s/\sqrt{\frac{\pi}{\pi-t}}\right), & \hbox{$0< t < 1/e$}
  \end{array}
\right.
\]
the discrete Gaussian constellation results in mutual information
\begin{equation}\label{eq:lattice-capacity}
I_D \geq \frac{1}{2}\log {(1+\SNR)} - \frac{6\varepsilon}{n}
\end{equation}
per channel use.
\end{ther}

\begin{rem}
It is easy to satisfy the condition $\frac{\pi\epsilon_t}{1-\epsilon_t}\leq \epsilon_{\Lambda}\left(\tilde{\sigma}_w\right)$ in Theorem \ref{theorem:capacity}. To see this, we note that
\[
\tilde{\sigma}_w = \frac{\sigma_s\sigma_w}{\sqrt{\sigma_s^2+\sigma_w^2}} < \sigma_w
\]
and that the flatness factor decreases fast with the standard deviation. Thus, the condition is basically $\sigma_w < \sigma_s$.
\end{rem}


Now we introduce the notion of constellations that are good for capacity, in the sense that the gap $\frac{6\varepsilon}{n}$ to the AWGN capacity is negligible. From (\ref{eq:lattice-capacity}) and the conditions of Theorem \ref{theorem:capacity}, we define

\begin{deft}[Good constellations in the sense of mutual information]
A lattice $\Lambda$ with a discrete Gaussian distribution is a good constellation for the AWGN channel if $\frac{\epsilon_{\Lambda}(\tilde{\sigma}_w)}{n}$ is negligible.
\end{deft}

\begin{rem}
Comparing with the definition $\epsilon_{\Lambda^{(n)}}(\sigma) \leq 2^{-\Omega(n)}$ of secrecy-good lattices\footnote{In fact, $\epsilon_{\Lambda^{(n)}}(\sigma) = o(\frac{1}{n})$ is enough to achieve strong secrecy, yet exponential vanishing is more desired.} \cite{LLBS_12}, we can see the condition of good constellations are less stringent. This is consistent with the known result that capacity-achieving codes can provide weak secrecy, but not strong secrecy \cite{BlochLaneman13}.
\end{rem}


\begin{rem}
Again, the statement of Theorem \ref{theorem:capacity} is non-asymptotical, i.e., it can hold even if $n=1$. The implication of (\ref{eq:lattice-capacity}) is that one may construct a capacity-achieving lattice code from a good constellation. In particular, one may choose a low-dimensional lattice with a small gap to the AWGN channel capacity as bounded in Theorem \ref{theorem:capacity}.
The construction will be addressed in a forthcoming paper. In the following section, we consider discrete Gaussian distribution over the entire lattice $L$ which is AWGN-good. The lattice $\Lambda$ in Theorem \ref{theorem:capacity} may or may not be the AWGN-good lattice $L$.
\end{rem}

\section{Lattice Gaussian Coding And Error Probability}

Now we describe the proposed coding scheme based on the lattice Gaussian distribution for the AWGN channel with power constraint $P$. The SNR is defined by $\SNR = P/\sigma_w^2$ for noise variance $\sigma_w^2$. Let $L$ be an AWGN-good lattice of dimension $n_L$. For the sake of generality, let the codebook be $L-\mathbf{c}$, where $\mathbf{c}$ is a proper shift as is often the case for various reasons in practice \cite{Forney00}. The encoder maps the information bits to points in $L-\mathbf{c}$, which obey the lattice Gaussian distribution $D_{L-\mathbf{c},\sigma_s}$:
\[
D_{L-\mathbf{c},\sigma_s}(\mathbf{x})= \frac{\frac{1}{(\sqrt{2\pi}\sigma_s)^{n_L}}e^{- \frac{\|\mathbf{x}\|^2}{2\sigma_s^2}}}{f_{\sigma_s,\mathbf{c}}(L)}, \quad \mathbf{x} \in L-\mathbf{c}.
\]
We assume the flatness factor is small, under certain conditions to be made precise in the following. Particularly, this means that the transmission power $P$ of this scheme tends to the variance $\sigma_s^2$.


Since the lattice points are not equally probable a priori in the lattice Gaussian
coding, we will use maximum-a-posteriori (MAP) decoding.
The following connection with MMSE was proven in \cite{LLBS_12} for the case $\mathbf{c}=0$.
For completeness, we give the proof for the general case.

\begin{prop}[Equivalence between MAP decoding and MMSE lattice decoding]\label{equivalence}
Let $\mathbf{x} \sim D_{L-\mathbf{c},\sigma_s}$ be the input signaling
of an AWGN channel where the noise variance is $\sigma_w^2$ per dimension. Then
MAP decoding is equivalent to Euclidean lattice decoding of~$L-\mathbf{c}$ using a scaling coefficient $\alpha=\frac{\sigma_s^2}{\sigma_s^2+\sigma_w^2}$, which is asymptotically equal to the MMSE coefficient $\frac{P}{P+\sigma_w^2}$ in the limit $\epsilon_{L}\left(\sigma_s/\sqrt{\frac{\pi}{\pi-t}}\right) \to 0$ for $0<t<\pi$.
\end{prop}

\begin{IEEEproof}
The received signal is given by $\mathbf{y}=\mathbf{x}+\mathbf{w}$, where $\mathbf{x} \in L-\mathbf{c}$ and $\mathbf{w}$ is the i.i.d. Gaussian noise vector of variance $\sigma_w^2$. Thus the MAP
decoding metric is given by
\begin{align*}
\mathbb{P}(\mathbf{x}|\mathbf{y}) &=
\frac{\mathbb{P}(\mathbf{x},\mathbf{y})}{\mathbb{P}(\mathbf{y})}
\propto
\mathbb{P}(\mathbf{y}|\mathbf{x})\mathbb{P}(\mathbf{x}) \\
&\propto
\exp\left(-\frac{\norm{\mathbf{y}-\mathbf{x}}^2}{2\sigma_w^2}-\frac{\norm{\mathbf{x}}^2}{2\sigma_s^2}
\right)\\
&\propto
\exp\left(-\frac{1}{2}\left(\frac{\sigma_s^2+\sigma_w^2}{\sigma_s^2\sigma_w^2}
\norm{\frac{\sigma_s^2}{\sigma_s^2+\sigma_w^2}\mathbf{y}-\mathbf{x}}^2
\right)\right).
\end{align*}
Therefore,
\begin{align}
\arg\max_{\mathbf{x}\in L-\mathbf{c}} \mathbb{P}(\mathbf{x}|\mathbf{y})
&= \arg\min_{\mathbf{x}\in L-\mathbf{c}}
\norm{\frac{\sigma_s^2}{\sigma_s^2+\sigma_w^2}\mathbf{y}-\mathbf{x}}^2 \nonumber\\
&= \arg\min_{\mathbf{x}\in L-\mathbf{c}}
\norm{\alpha{\mathbf{y}}-\mathbf{x}}^2 \label{eq:renormalized-metric}
\end{align}
where $\alpha=\frac{\sigma_s^2}{\sigma_s^2+\sigma_w^2}$ is known, thanks to Lemma \ref{Micciancio_Regev_Lemma43}, to be asymptotically equal to the MMSE coefficient $\frac{P}{P+\sigma_w^2}$.
\end{IEEEproof}

Therefore, the MAP decoder is simply given by
\begin{equation}\label{MAPdecoder}
\hat{\mathbf{x}}=Q_{L-\mathbf{c}}\left(\alpha{\mathbf{y}}\right)
\end{equation}
where $Q_{L-\mathbf{c}}$ denotes, in a similar fashion to $Q_{L}$, the minimum Euclidean-distance decoder for shifted lattice $L-\mathbf{c}$.

\subsection{Error Probability}

Now let us analyze the average error probability of the MAP decoder. In Appendix \ref{appendixII}, we derive the following lemma which shows that the error probability of the proposed scheme admits almost the same expression as that of Poltyrev \cite{Poltyrev94}, with $\sigma_w^2$ replaced by $\tilde{\sigma}_w^2$ (recall $\tilde{\sigma}_w = \frac{\sigma_s\sigma_w}{\sqrt{\sigma_s^2+\sigma_w^2}}$).

\begin{lem}\label{lem:error-prob-MMSE}
For any lattice $L$, the average error probability of the MAP decoder (\ref{MAPdecoder}) for a lattice codebook of distribution $D_{L-\mathbf{c},\sigma_s}$ is bounded by
\begin{equation}\label{Pe-Poltyrev-MMSE}
P_e \in \left[\frac{1 - \epsilon_{L}\left(\frac{\sigma_s^2}{\sqrt{\sigma_s^2+\sigma_w^2}}\right)}{1+ \epsilon_{L}\left(\sigma_s\right)},\frac{1 + \epsilon_{L}\left(\frac{\sigma_s^2}{\sqrt{\sigma_s^2+\sigma_w^2}}\right)}{1- \epsilon_{L}\left(\sigma_s\right)}\right] P_e(L,\tilde{\sigma}_w^2)
\end{equation}
where
\[
P_e(L,\tilde{\sigma}_w^2) = \frac{1}{\left(\sqrt{2\pi}\tilde{\sigma}_w\right)^{n_L}} \int_{\mathcal{\overline{V}}(L)}
\exp\left\{- \frac{\|\mathbf{y}\|^2}{2\tilde{\sigma}_w^2}\right\} d\mathbf{y}
\]
is the error probability of infinite lattice decoding for noise variance $\tilde{\sigma}_w^2$.

\end{lem}

By the well-known result of Poltyrev \cite{Poltyrev94}, if $L$ is AWGN-good, then the error probability of infinite lattice coding for noise variance $\sigma_w^2$ is asymptotically bounded by
\begin{equation}\label{Pe-Poltyrev1}
P_e(L,\tilde{\sigma}_w^2) \leq e^{-n_L E_P(\gamma_{L}({\sigma_w}))}
\end{equation}
where $E_P(\mu)$ denotes the Poltyrev exponent
\begin{equation}
E_P(\mu)=\begin{cases} \frac{1}{2} \left[(\mu-1)-\log \mu\right] & 1 < \mu \leq 2\\
\frac{1}{2} \log \frac{e\mu}{4} & 2 \leq \mu \leq 4\\
\frac{\mu}{8} & \mu \geq 4.
\end{cases}
\end{equation}

Consequently, we have the following lemma for error performance of AWGN-good lattices.

\begin{lem}\label{lem:error-prob}
If $L$ is AWGN-good, then the average error probability of the MAP decoder (\ref{MAPdecoder}) is bounded by
\begin{equation}\label{Pe-Poltyrev}
P_e \leq \frac{1 + \epsilon_{L}\left(\frac{\sigma_s^2}{\sqrt{\sigma_s^2+\sigma_w^2}}\right)}{1- \epsilon_{L}\left(\sigma_s\right)} e^{-n_L E_P(\gamma_{L}(\tilde{\sigma}_w))}
\end{equation}
\end{lem}

If $L$ is good for AWGN, $P_e$ will vanish if $\gamma_{L}(\tilde{\sigma}_w) > 2\pi e$, i.e.,
\begin{equation}\label{eq:condition1}
V(L)^{2/{n_L}} > 2\pi e \tilde{\sigma}_w^2.
\end{equation}
In (\ref{Pe-Poltyrev}), we also need to make  $\epsilon_{L}\left(\frac{\sigma_s^2}{\sqrt{\sigma_s^2+\sigma_w^2}}\right) \rightarrow 0$ and $\epsilon_{L}\left(\sigma_s\right)\rightarrow0$ so that $P_e$ approaches the Poltyrev bound. Obviously, the first condition subsumes the second one. So, for mod-$p$ lattices, we can satisfy it by making
\begin{equation}\label{eq:Gauss-condition}
\gamma_{L}\left(\frac{\sigma_s^2}{\sqrt{\sigma_s^2+\sigma_w^2}}\right)=\frac{V(L)^{2/{n_L}}}{2\pi\frac{\sigma_s^4}{{\sigma_s^2+\sigma_w^2}} }<1,
\end{equation}

It is worth pointing out that the AWGN-goodness of $L$ and the flatness condition $\epsilon_{L}\left(\frac{\sigma_s^2}{\sqrt{\sigma_s^2+\sigma_w^2}}\right) \rightarrow 0$ are not contradictory, since they involve different variances (i.e., $\tilde{\sigma}_w^2$ and $\frac{\sigma_s^4}{{\sigma_s^2+\sigma_w^2}}$ whose ratio is essentially the SNR). In fact, conditions (\ref{eq:condition1}) and (\ref{eq:Gauss-condition}) are compatible if
\begin{equation}\label{eq:condition2}
\frac{\sigma_s^2}{\sigma_w^2} > e
\end{equation}
which is a very mild condition, i.e, the SNR is larger than $e$.

\subsection{Rate}

Now, to satisfy the volume constraint (\ref{eq:condition1}), we choose the fundamental volume $V(L)$ such that
\begin{equation}\label{eq:condition3}
V(L)^{2/{n_L}} = 2\pi e \tilde{\sigma}_w^2 (1 + \varepsilon'')
\end{equation}
for some small $\varepsilon'' \to 0$.

By Lemma \ref{Micciancio_Regev_Lemma43}, we have
\begin{equation}\label{eq:condition4}
\sigma_s^2 \geq \frac{1}{1+\frac{2\pi \varepsilon_t}{n_L(1-\varepsilon)}}P,
\end{equation}
where $\varepsilon = \epsilon_{\Lambda}\left(\sigma_s/\sqrt{\frac{\pi}{\pi-t}}\right) < 1$ and $\varepsilon_t$ is as defined in Lemma \ref{Micciancio_Regev_Lemma43}.

The rate $R$ of the code can be as large as the entropy rate of $\mathbf{x}$. By Lemma \ref{proposition4} and (\ref{eq:condition3}), the maximum rate $R_{\max}$ is bounded from below by
\begin{align}
R_{\max} &\geq \log (\sqrt{2 \pi e}\sigma_s) - \frac{1}{{n_L}}\log {V(L)} - \varepsilon'\nonumber\\
&= \log (\sqrt{2 \pi e}\sigma_s) - \frac{1}{2}\log \left({2\pi e \frac{\sigma_s^2\sigma_w^2}{\sigma_s^2+\sigma_w^2}}\right) - \nonumber\\
&\quad \quad \quad \frac{1}{2}\log(1+\varepsilon'') - \varepsilon'\nonumber\\
&\geq \frac{1}{2}\log {\left(1+\frac{\sigma_s^2}{\sigma_w^2}\right)} - \frac{1}{2}\varepsilon'' - \varepsilon' \nonumber
\end{align}
where $\varepsilon'= - \frac{\log(1-\varepsilon)}{n_L} + \frac{\pi\varepsilon_t}{n_L(1-\varepsilon)}$.
Thus, applying (\ref{eq:condition4}), we obtain
\begin{align}
R_{\max} &\geq \frac{1}{2}\log {\left(1+\SNR\right)} - \frac{\pi \varepsilon_t}{n_L(1-\varepsilon)} - \frac{1}{2}\varepsilon'' - \varepsilon' \label{rate}\\
& \to \frac{1}{2}\log {(1+\SNR)} \nonumber
\end{align}
if $\varepsilon \to 0$ and $\varepsilon'' \to 0$. It can be verified that (\ref{eq:condition1}) and $\varepsilon \to 0$ are compatible for mod-$p$ lattices if
\begin{equation}\label{eq:condition22}
\frac{\sigma_s^2}{\sigma_w^2} > \frac{\pi}{\pi-t}e-1.
\end{equation}
For $t \to 0$, the required SNR is larger than $e-1$.

Therefore, using this lattice Gaussian codebook, we can achieve a rate arbitrarily close to the channel capacity while making the error probability vanish exponentially, as long as $\SNR>e$ (cf. conditions (\ref{eq:condition2}) and (\ref{eq:condition22})). We summarize the results in the following theorem:

\begin{ther}[Coding theorem for lattice Gaussian coding]\label{theorem3}
Consider a lattice code whose codewords are drawn from the discrete Gaussian distribution $D_{L-\mathbf{c},\sigma_s}$ for an AWGN-good lattice $L$. If $\SNR>e$, then any rate (\ref{rate}) up to the channel capacity $\frac{1}{2}\log {(1+\SNR)}$ is achievable, while the error probability of MMSE lattice decoding vanishes exponentially fast as in (\ref{Pe-Poltyrev}).
\end{ther}


\begin{rem}
Given the rate $R < C=\frac{1}{2}\log {(1+\SNR)}$, we can make the VNR more explicit. Since $\sigma_s^2 \geq P$, we have
\begin{equation}\label{eq:bound23}
  \frac{\sigma_s^2+\sigma_w^2}{\sigma_w^2} \geq e^{2C}.
\end{equation}
From Lemma \ref{proposition4}, we derive
\begin{equation}\label{eq:bound24}
  \frac{2\pi e \sigma_s^2}{V(\Lambda)^{2/n}}  \leq 2^{2R} e^{2\varepsilon'}.
\end{equation}
Dividing (\ref{eq:bound23}) by (\ref{eq:bound24}), we obtain
\[
{\gamma_{L}(\tilde{\sigma}_w)} \geq 2 \pi e \cdot e^{2(C-R)}\frac{e^{-2\varepsilon'}}{1+\frac{2\pi \varepsilon_t}{n(1-\varepsilon_t)}}.
\]
Consequently, the error probability can be bounded by (\ref{Pe-Poltyrev}) with the VNR given above.
\end{rem}

\subsection{Comparison with Voronoi Constellations}

Now we compare with Voronoi constellations or nested lattice codes where the shaping lattice is good for quantization \cite{ErezZamir04}. In such a scheme, the transmitted signal (subject to a random dither) is uniformly distributed on the Voronoi region of the shaping lattice. It is shown in \cite{Zamir_Feder_96} that such a uniform distribution converges to a Gaussian distribution in a weak sense, that is, the \emph{normalized} Kullback-Leibler divergence (i.e., divided by the dimension) tends to zero. Since the Voronoi region of a quantization-good lattice converges to a sphere, the peak power is $n_L P$ asymptotically for average power $P$.

Our proposed scheme uses a discrete Gaussian distribution over $L$, hence requiring neither shaping nor dithering. Since it uses the entire lattice, the peak power seems to be infinite. Nevertheless, this need not be the case. By the generalized AEP, if $\mathbf{x} \sim D_{L-\mathbf{c},\sigma_s}$, we have
\begin{equation}\label{prob-sphere2}
\mathbb{P}(\|\mathbf{x}\|>\rho \cdot \sqrt{n_L}\sigma_s) <
\frac{1+\epsilon_L(\sigma_s)}{1-\epsilon_L(\sigma_s)} e^{-n_L E_{\mathrm{sp}}(\rho^2)}.
\end{equation}
As long as $\epsilon_L(\sigma_s)$ is bounded by a constant, the right-hand side of (\ref{prob-sphere2}) goes to zero for any $\rho > 1$. Therefore, in practice, the outer points need not to be sent, and the constellation  points can be drawn from a sphere of radius arbitrarily close to $\sqrt{n_L}\sigma_s$. The peak power can be arbitrarily close to $n_L P$, which is the same as that of the Voronoi constellation.
Thus, in this aspect, the lattice Gaussian codebook is very similar to a finite constellation.

It is also interesting to note that MMSE lattice decoding of outer points (i.e., those of large norm $\|\mathbf{x}\|^2$) is very likely to fail, since the equivalent noise $\left(\alpha-1\right)\mathbf{x}+\alpha\mathbf{w}$ will be very strong in this case. Nevertheless, the average error probability still admits almost the same expression as Poltyrev's (with $\sigma_w^2$ replaced by $\tilde{\sigma}_w^2$). This is because outer points are sent with a small probability, thus carrying little weight in the average error probability.

The error analysis of Erez and Zamir's scheme is somewhat involved, since the equivalent noise in \cite{ErezZamir04} is not Gaussian. Yet they also proved their scheme has almost the same error performance as Poltyrev's (with $\sigma_w^2$ replaced by $\tilde{\sigma}_w^2$), hence almost the same as our proposed scheme.

\section{Implementation}

The afore-going analysis shows that the problem of achieving the AWGN channel capacity is reduced to that of finding an AWGN-good lattice for noise variance $\tilde{\sigma}_w^2$, i.e., a lattice $L$ whose error probability $P_e(L,\tilde{\sigma}_w^2)\to 0$ as long as $\gamma_{L}(\tilde{\sigma}_w)<2\pi e$.
Forney {\it et al.} gave constructions of such lattices in \cite{Forney00}.
We focus on Construction A. Let $\Lambda_1/\Lambda_2$ be a lattice partition where $\Lambda_1$ is the fine lattice and $\Lambda_2$ is the coarse lattice, both of dimension $n$. It is worth pointing out that $\Lambda_1$ and $\Lambda_2$ can be simple low-dimensional lattices such as $\mathbb{Z}$ and $2\mathbb{Z}$. Let $\mathcal{C} \in \Lambda_1^N/\Lambda_2^N$ be a linear code of length $N$. Construction A of $L$ is given by
\[
L = \{\mathbf{x} = \mathbf{a} + \mathbf{b}  | \mathbf{a} \in \mathcal{C}, \mathbf{b} \in \Lambda_2^N\}.
\]
Thus the dimension of $L$ is $n_L=nN$. If $\Lambda_1/\Lambda_2 \cong (\mathbb{Z}_p)^r$ with $p$ prime, then linear codes over $\mathrm{GF}(p^r)$ may be used. The case of $r=1$ corresponds to the usual mod-$p$ lattices. More generally, mod-$q$ lattices where $q$ is not necessarily a prime can be used, and the corresponding code is defined over a ring.

\begin{figure*}[t]

\centering\centerline{\epsfig{figure=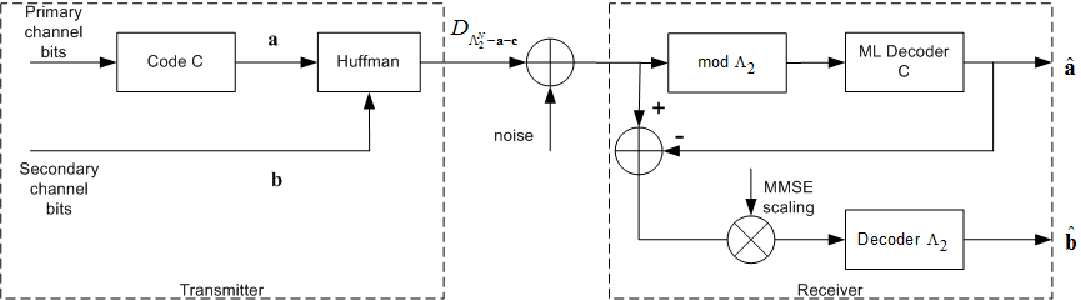,width=16cm}}

\caption{Block diagram of encoding and decoding of the proposed lattice Gaussian coding scheme.}

\vspace{-0.5cm}

\label{fig:block-diagram}
\end{figure*}

In this Section, we describe the implementation of the proposed lattice Gaussian coding scheme for Construction A. In general, we need shaping over the code $\mathcal{C}$, since the cosets are not necessarily equally probable. Yet, by the first part of Lemma \ref{lemma6}, if the flatness factor $\epsilon_{\Lambda_2}(\sigma_s)$ of the coarse lattice is sufficiently small, the cosets are nearly equally probable. In this case, shaping over the code may be dropped, and the implementation of the scheme will be greatly simplified. We now present such an encoding procedure which produces codewords from a distribution close to $D_{L-\mathbf{c},\sigma_s}$. The overall block diagram of the encoder and decoder is given in Fig. \ref{fig:block-diagram}.

\subsection{Encoding Procedure}

The procedure is comprised of two steps:
\begin{enumerate}
  \item Generate a codeword $\mathbf{a}$ of code $\mathcal{C} \in \Lambda_1^N/\Lambda_2^N$ from a uniform distribution of the input bits;
  \item Generate a point $\mathbf{x}\in\Lambda_2^N-\mathbf{a}-\mathbf{c}$ from distribution $D_{\Lambda_2^N-\mathbf{a}-\mathbf{c},\sigma_s}$.
\end{enumerate}
Similar procedures have been used before \cite{Forney_Wei_89,Kschischang_Pasupathy,Palgy12}. Step 1 is the usual block coding of $\mathcal{C}$ at a fixed rate of input bits, which are referred to as the primary channel bits. Step 2 has a variable rate due to the secondary channel bits.

We will show that the resulting distribution of codeword $\mathbf{x}$ is very close to $D_{L-\mathbf{c},\sigma_s}$ if $\epsilon_{\Lambda_2}(\sigma_s)$ is small. Let $\mathbf{a} = [\mathbf{a}_1^T, \mathbf{a}_2^T, \ldots \mathbf{a}_N^T]^T$ and
$\mathbf{c} = [\mathbf{c}_1^T, \mathbf{c}_2^T, \ldots \mathbf{c}_N^T]^T$ be the decompositions into $N$ elements.

Note that Step 2 consists of $N$ independent realizations of $D_{\Lambda_2-\mathbf{a}_i-\mathbf{c}_i,\sigma_s}$, which gives rise to distribution $D_{\Lambda_2^N-\mathbf{a}-\mathbf{c},\sigma_s}$ exactly. To realize $D_{\Lambda_2-\mathbf{a}_i-\mathbf{c}_i,\sigma_s}$, we use Huffman coding to construct a source code for distribution $D_{\Lambda_2-\mathbf{a}_i-\mathbf{c}_i,\sigma_s}$ over each coset. This has already been implemented in \cite{Kschischang_Pasupathy,Palgy12}. Basically, one may use Huffman coding to construct a code tree for distribution $\Lambda_2-\mathbf{a}_i-\mathbf{c}_i$. This is quite affordable since $\Lambda_2$ is a simple low-dimension lattice such as $\mathbb{Z}$ or $\mathbb{Z}^2$. To map information bits to lattice points, one just applies Huffman decoding: traverse the tree until reaching a leave (i.e., a lattice point).



By Lemma \ref{lemma5}, the flatness factor of $\Lambda_2^N$ is given by
\begin{equation}\label{eq:flatness-2N}
\epsilon_{\Lambda_2^N}(\sigma_s) = [1+\epsilon_{\Lambda_2}(\sigma_s)]^N-1 \approx N\epsilon_{\Lambda_2}(\sigma_s)
\end{equation}
if $\epsilon_{\Lambda_2}(\sigma_s)$ is small. Although the flatness factor increases with $N$, we can keep it under control by making $\epsilon_{\Lambda_2}(\sigma_s)$ sufficiently small.

Then, we invoke the second part of Lemma \ref{lemma6} to show that the variational distance between the resultant distribution $D'_{L-\mathbf{c},\sigma_s}$ of $\mathbf{x}$ and $D_{L-\mathbf{c},\sigma_s}$ is bounded as
\begin{equation}\label{eq:distanceV2D}
\V(D'_{L-\mathbf{c},\sigma_s},D_{L-\mathbf{c},\sigma_s}) \leq \frac{2\epsilon_{\Lambda_2^N}(\sigma_s)}{1-\epsilon_{\Lambda_2^N}(\sigma_s)} \approx {2N\epsilon_{\Lambda_2}(\sigma_s)}.
\end{equation}
Therefore, the resultant distribution is very close to $D_{L-\mathbf{c},\sigma_s}$. We can  make $\epsilon_{\Lambda_2}(\sigma_s)$ negligible if there is enough power $\sigma_s^2$. This is not hard to achieve since $\epsilon_{\Lambda_2}(\sigma_s)$ decreases faster with $\sigma_s^2$.

\begin{figure}[t]

\centering\centerline{\epsfig{figure=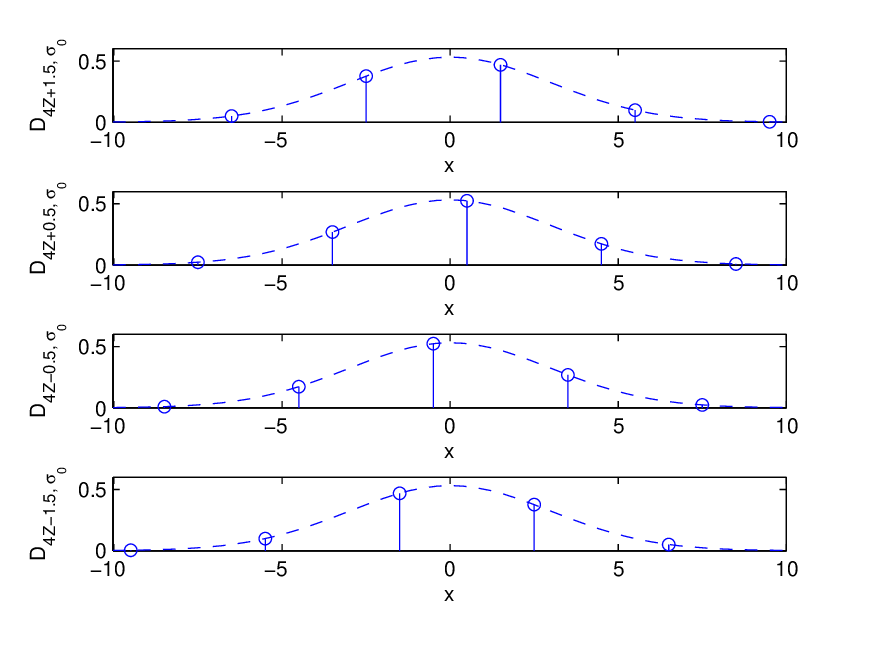,width=9cm}}

\caption{Lattice Gaussian distribution (circle) over four cosets $4\mathbb{Z}+1.5$, $4\mathbb{Z}+0.5$, $4\mathbb{Z}-0.5$ and $4\mathbb{Z}-1.5$ of $4\mathbb{Z}$ for $\sigma_s=3$ and $c=0.5$. The profile (dashed) is the underlying continuous Gaussian distribution.}

\vspace{-0.5cm}

\label{fig:Gaussian-4Z}
\end{figure}

\begin{exap}
For mod-$q$ lattices $L = \{\mathbf{x} = \mathbf{a} + \mathbf{b}  | \mathbf{a} \in \mathcal{C}, \mathbf{b} \in q\mathbb{Z}^N\}$, we only need to handle the one-dimensional distribution
$D_{q\mathbb{Z}-{a}_i-{c}_i,\sigma_s}$ over a coset of $q\mathbb{Z}$. Fig. \ref{fig:Gaussian-4Z} gives an example of $4\mathbb{Z}$ where the shift $c=0.5$ and $\sigma_s=3$. In this case, since the corresponding flatness factor $\epsilon_{4\mathbb{Z}}(\sigma_s)=3\times 10^{-5}$, the four cosets of $4\mathbb{Z}$ are essentially equally probable.
\end{exap}

\begin{rem}
It is straightforward to extend the encoding procedure to multilevel lattices \cite{Forney00}. In this case, the variational distance bound (\ref{eq:distanceV2D}) is determined by the flatness factor of the bottom lattice.
\end{rem}

\subsection{Construction A from Binary Codes}

\begin{figure}[t]

\centering\centerline{\epsfig{figure=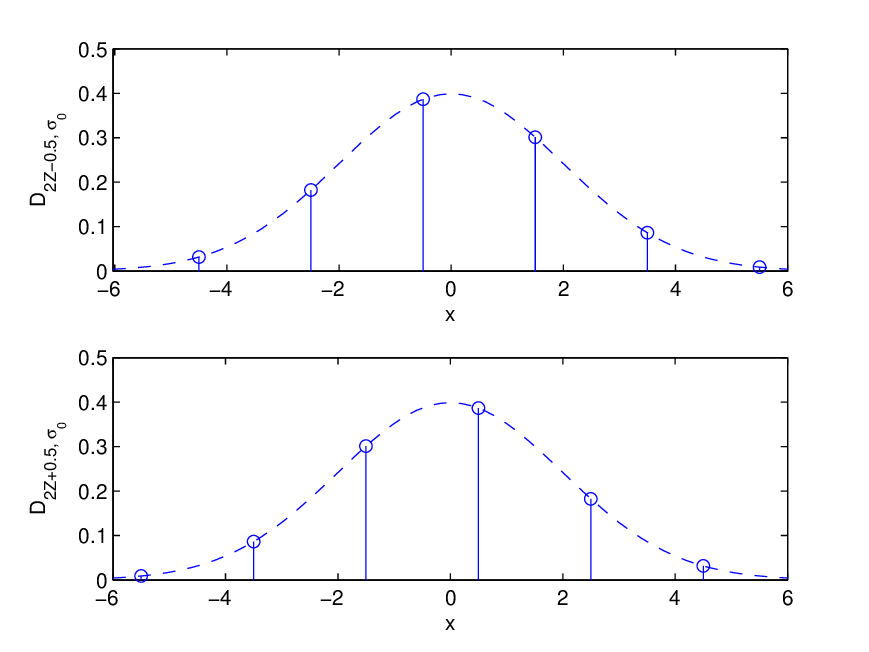,width=9cm}}

\caption{Lattice Gaussian distribution (circle) over two cosets $2\mathbb{Z}-\frac{1}{2}$ and $2\mathbb{Z}+\frac{1}{2}$ for $\sigma_s=2$. The profile (dashed) is the underlying continuous Gaussian distribution.}

\vspace{-0.5cm}

\label{fig:Gaussian-2Z}
\end{figure}

In some cases, it is possible for this procedure to produce the exact distribution $D_{L-\mathbf{c},\sigma_s}$. We give an example for the standard Construction A from binary codes, i.e., $L=2\mathbb{Z}^{N}+\mathcal{C}$ where $\mathcal{C}$ is a binary code over $\mathrm{GF}(2)$. In practice, the lattice $L$ is often shifted by $\mathbf{c}=\frac{1}{2}\mathbf{1}^T$. Due to the symmetry of this lattice constellation, all the codewords of $\mathcal{C}$ (after the shift) have the same Euclidean norm (i.e., each component of its codewords is $\pm \frac{1}{2}$). Thus, all the cosets have the same probability, and accordingly, the codewords of $\mathcal{C}$ are indeed uniformly distributed. We only need to implement the encoding for one-dimensional distributions $D_{2\mathbb{Z}- \frac{1}{2},\sigma_s}$ and $D_{2\mathbb{Z}+ \frac{1}{2},\sigma_s}$, which is easy. Fig. \ref{fig:Gaussian-2Z} shows distributions $D_{2\mathbb{Z}- \frac{1}{2},\sigma_s}$ and $D_{2\mathbb{Z}+ \frac{1}{2},\sigma_s}$. Due to the symmetry, the two cosets are equally probable, regardless of the value of $\sigma_s$.

\begin{exap}
The checkerboard lattice $D_n$ can be constructed from the $(n,n-1,2)$ binary parity-check code:
\[
D_n = 2\mathbb{Z}^n + (n,n-1,2).
\]
The Gosset lattice $E_8$ can be constructed from the extended $(7,4)$ Hamming code:
\[
E_8 = 2\mathbb{Z}^8 + (8,4,4).
\]
For such lattices as well as trellis codes constructed from binary convolutional codes $\mathcal{C}$ \cite{Forney1}, the implementation of lattice Gaussian coding is convenient.
\end{exap}

\subsection{Decoding}

The decoding also benefits from the proposed encoding procedure. Following \cite{Forney00}, we use stage-by-stage decoding as shown in Fig. \ref{fig:block-diagram}. The first stage is to decode the code $\mathcal{C}$ on the mod-$\Lambda_2$ channel. Since the cosets are uniformly distributed in the proposed encoding procedure, this is just the standard maximum-likelihood (ML) decoding. Then, the codeword is subtracted out, and MAP decoding is applied to $\Lambda_2$ (which is equivalent to MMSE lattice decoding).

\section{Conclusions and Discussion}

In this paper, we have proved that the lattice Gaussian distribution over an AWGN-good lattice can achieve the $\frac{1}{2}\log (1+\SNR)$ capacity under MMSE lattice decoding. The crucial technique of the proof is the flatness factor, which enables us to show the error probability admits almost the same form as that of Poltyrev's infinite lattice coding if $\SNR > e$.
Regarding the implementation of lattice Gaussian shaping, we have derived a bound on the variational distance between $D_{L-\mathbf{c},\sigma_s}$ and the distribution resulting from the intuitive method where shaping is only applied to the bottom lattice $\Lambda_2$; this bound is almost zero if $\epsilon_{\Lambda_2}({\sigma_s})$ is negligible. Again, it is worth mentioning that the conditions on the flatness factor do not have to be asymptotic. In general, these are mild conditions, which can be met either by scaling down the component lattices or by moderately increasing the signal power.

Finally, we note adding dither to lattice Gaussian shaping \cite{Palgy12} has a similar effect as the flatness factor, in the sense that badly positioned constellations are avoided and the averaging behavior is constantly obtained.

\section*{Acknowledgments}
The authors would like to thank Damien Stehl\'e, Laura Luzzi, Ram Zamir, Ashish Khisti, Shlomo Shamai and Daniel Dadush for helpful discussions.

\appendices

\section{Proof of Lemma \ref{lem:sphere}} \label{appendix0}


Denote by $\mathcal{B}_n$ the $n$-dimensional unit ball. Since $\mathbf{x} \sim D_{\Lambda,\sigma,\mathbf{c}}$, we have
\begin{equation}
\p(\|\mathbf{x}-\mathbf{c}\|>\rho \sqrt{n}\sigma) = \frac{f_{\sigma}((\Lambda-\mathbf{c}) \setminus \rho \sqrt{n}\sigma\mathcal{B}_n)}{f_{\sigma}(\Lambda-\mathbf{c})}.
\end{equation}

By the definition of the flatness factor, we have
\[
f_{\sigma}(\Lambda-\mathbf{c})={f_{\sigma,{\bf c}}(\Lambda)} \geq \frac{1- \epsilon_{\Lambda}(\sigma)}{1+ \epsilon_{\Lambda}(\sigma)} f_{\sigma}(\Lambda).
\]
By \cite[Lemma 1.5]{Banaszczyk}, for any $\rho > 1$ we have\footnote{In \cite[Lemma 1.5]{Banaszczyk}, there is another factor 2 on the right-hand side of the bound, yet it can be removed after a careful check.}
\[
f_{\sigma}((\Lambda-\mathbf{c}) \setminus \rho\sqrt{n}\sigma\mathcal{B}_n) < \left(\rho\cdot e^{(1-\rho^2)/2} \right)^n f_{\sigma}(\Lambda).
\]
Note that $\rho\cdot e^{(1-\rho^2)/2}<1$ for $\rho>1$.
Thus,
\begin{align*}
\p(\|\mathbf{x}-\mathbf{c}\|>\rho \sqrt{n}\sigma) &\leq \frac{1+ \epsilon_{\Lambda}(\sigma)}{1- \epsilon_{\Lambda}(\sigma)}\left(\rho\cdot e^{(1-\rho^2)/2} \right)^n \\
& \leq \frac{1+ \epsilon_{\Lambda}(\sigma)}{1- \epsilon_{\Lambda}(\sigma)} \cdot e^{-nE_{\mathrm{sp}}(\rho^2)}
\end{align*}
using the definition of the sphere-packing exponent.

%
%

\section{Proof of Theorem \ref{theorem:capacity}} \label{appendixI}

Let $\Phi$ denote a continuous Gaussian random vector of zero mean
and variance $\sigma_s^2$ per dimension, and write $\Y=\mathsf{X}+\mathsf{W}$ and
$\Y'=\Phi+\mathsf{W}$, respectively. Obviously, $\Y'$ is a continuous Gaussian random vector of zero mean
and variance $\sigma_s^2+\sigma_w^2$ per dimension. The difference between the
mutual information achieved by $\mathsf{X}$ and by $\Phi$ is given
by
\begin{align*}
\I(\Phi;\Y') - \I(\mathsf{X};\Y) &= h(\Y') -h(\Y)
\end{align*}
where $h(\cdot)$ is the differential entropy. We note that the Kullback-Leibler divergence can be rewritten as
\begin{align*}
\D(\Y\|\Y') &= -\int_{\R^n} p_{\Y}(\mathbf{y}) \log
{p_{\Y'}(\mathbf{y})} d\mathbf{y} - h(\Y)\\
&\stackrel{(a)}{=} \int_{\R^n}   \frac{\|\mathbf{y}\|^2p_{\Y}(\mathbf{y})}{2(\sigma_s^2+\sigma_w^2)} d\mathbf{y} + n \log \sqrt{2\pi(\sigma_s^2+\sigma_w^2)} - h(\Y)\\
&\stackrel{(b)}{\geq} \frac{n\sigma_s^2-\frac{2\pi\epsilon_t}{1-\epsilon_t}\sigma_s^2 +n\sigma_w^2}{2(\sigma_s^2+\sigma_w^2)} \\&\quad \quad \quad + n \log \sqrt{2\pi(\sigma_s^2+\sigma_w^2)} - h(\Y)\\
&\geq \frac{n}{2} - \frac{\pi\epsilon_t}{1-\epsilon_t} + n \log \sqrt{2\pi(\sigma_s^2+\sigma_w^2)} - h(\Y)\\
&= n\log\sqrt{2\pi e(\sigma_s^2+\sigma_w^2)} - h(\Y) - \frac{\pi\epsilon_t}{1-\epsilon_t}\\
&= h(\Y')- h(\Y) - \frac{\pi\epsilon_t}{1-\epsilon_t},
\end{align*}
where (a) is obtained by expanding the Gaussian density $p_{\Y'}(\mathbf{y})$, and (b) is due to the fact that the second moment of $\Y$ equals the sum of the second moment of $\mathsf{X}$ (Lemma~\ref{Micciancio_Regev_Lemma43}) and the variance of $\mathsf{W}$.
Therefore, we have
\begin{align*}
\I(\Phi;\Y') - \I(\mathsf{X};\Y) &\leq \D(\Y\|\Y') + \frac{\pi\epsilon_t}{1-\epsilon_t} \\
&\stackrel{(a)}{\leq} \log (1+4\varepsilon) + \frac{\pi\epsilon_t}{1-\epsilon_t}\\
&{\leq} 4\varepsilon + \frac{\pi\epsilon_t}{1-\epsilon_t}\\
&\stackrel{(b)}{\leq} 5\varepsilon
\end{align*}
where (a) and (b) are due to Lemma \ref{corol-KL} and the condition $\frac{\pi\epsilon_t}{1-\epsilon_t}\leq \varepsilon$, respectively.


Since the continuous Gaussian
distribution achieves capacity $\I(\Phi;\Y')=\frac{n}{2}\log\left(1+\frac{\sigma_s^2}{\sigma_w^2}\right)$, we have
\begin{equation}\label{eq:lattice-capacity1}
I_D \geq \frac{1}{2}\log\left(1+\frac{\sigma_s^2}{\sigma_w^2}\right) - \frac{5\varepsilon}{n}
\end{equation}
per channel use.

It remains to bound $\sigma_s^2$ by $P$. By lemma \ref{Micciancio_Regev_Lemma43}, we have
\begin{equation}\label{eq:condition44}
\sigma_s^2 \geq \frac{1}{1+\frac{2\pi \epsilon_t}{n(1-\epsilon_t)}}P.
\end{equation}
This leads to
\begin{align}
\frac{1}{2}\log {\left(1+\frac{\sigma_s^2}{\sigma_w^2}\right)} &\geq  \frac{1}{2}\log {\left(1+\frac{\SNR}{1+\frac{2\pi \epsilon_t}{n(1-\epsilon_t)}}\right)}\nonumber\\
&\geq \frac{1}{2}\log {(1+\SNR)} - \frac{1}{2} \log \left(1+\frac{2\pi \epsilon_t}{n(1-\epsilon_t)}\right) \label{eq:bound-sigma0-sigma} \\
&\geq \frac{1}{2}\log {(1+\SNR)} - \frac{\pi \epsilon_t}{n(1-\epsilon_t)} \nonumber\\
&\geq \frac{1}{2}\log {(1+\SNR)} - \frac{\varepsilon}{n} \nonumber
\end{align}
where the last step is again due to the condition $\frac{\pi\epsilon_t}{1-\epsilon_t}\leq \varepsilon$.
The theorem is proven by combining this with (\ref{eq:lattice-capacity1}).

\section{proof of Lemma \ref{lem:error-prob} } \label{appendixII}

Suppose $\mathbf{x} \in L-\mathbf{c}$ is sent. The received signal after MMSE scaling can be written as
\begin{equation}\label{eq:equiv-noise}
\mathbf{y} = \alpha (\mathbf{x}+\mathbf{w})= \mathbf{x}+\left(\alpha-1\right)\mathbf{x}+\alpha\mathbf{w}.
\end{equation}

The decoding error probability associated with $\mathbf{x}$ is given by
\begin{align}
P_e(\mathbf{x}) &= 1-\int_{\mathbf{x}+\mathcal{{V}}(L)}
{\frac{1}{(\sqrt{2\pi}\alpha\sigma)^{n_L}}\exp\left\{- \frac{\|\mathbf{y}-\alpha\mathbf{x}\|^2}{2\alpha^2\sigma_w^2}\right\}} d\mathbf{y} \nonumber\\
&= 1- \int_{\mathcal{{V}}(L)}
{\frac{1}{(\sqrt{2\pi}\alpha\sigma)^{n_L}}\exp\left\{- \frac{\|\mathbf{y}-(\alpha-1)\mathbf{x}\|^2}{2\alpha^2\sigma_w^2}\right\}} d\mathbf{y} \nonumber\\
&= \int_{\mathcal{\overline{V}}(L)}
{\frac{1}{(\sqrt{2\pi}\alpha\sigma)^{n_L}}\exp\left\{- \frac{\|\mathbf{y}-(\alpha-1)\mathbf{x}\|^2}{2\alpha^2\sigma_w^2}\right\}} d\mathbf{y} \nonumber
\end{align}
where $\mathcal{\overline{V}}(L)$ denotes the complement of the Voronoi region $\mathcal{{V}}(L)$ in $\mathbb{R}^{n_L}$.

The average decoding probability is given by
\begin{align}
P_e &= \sum_{\mathbf{x}\in L-\mathbf{c}}{\frac{\frac{1}{(\sqrt{2\pi}\sigma_s)^{n_L}}e^{- \frac{\|\mathbf{x}\|^2}{2\sigma_s^2}}}{f_{\sigma_s,\mathbf{c}}(L)}} P_e(\mathbf{x}) \nonumber\\
&= \sum_{\mathbf{x}\in L-\mathbf{c}}{\frac{\frac{1}{(\sqrt{2\pi}\sigma_s)^{n_L}}e^{- \frac{\|\mathbf{x}\|^2}{2\sigma_s^2}}}{f_{\sigma_s,\mathbf{c}}(L)}} \times \nonumber\\
&\quad \quad \quad \int_{\mathcal{\overline{V}}(L)}
{\frac{1}{(\sqrt{2\pi}\alpha\sigma)^{n_L}}\exp\left\{- \frac{\|\mathbf{y}-(\alpha-1)\mathbf{x}\|^2}{2\alpha^2\sigma_w^2}\right\}} d\mathbf{y} \nonumber\\
&= \frac{\frac{1}{({2\pi}\alpha\sigma_s\sigma)^{n_L}}}{f_{\sigma_s,\mathbf{c}}(L)} \times \nonumber\\
&\quad \quad \quad \sum_{\mathbf{x}\in L-\mathbf{c}}{} \int_{\mathcal{\overline{V}}(L)}
{e^{- \frac{\|\mathbf{x}\|^2}{2\sigma_s^2}}\exp\left\{- \frac{\|\mathbf{y}-(\alpha-1)\mathbf{x}\|^2}{2\alpha^2\sigma_w^2}\right\}} d\mathbf{y}
\nonumber\\
&= \frac{\frac{1}{({2\pi}\alpha\sigma_s\sigma)^{n_L}}}{f_{\sigma_s,\mathbf{c}}(L)} \sum_{\mathbf{x}\in L-\mathbf{c}}{} \int_{\mathcal{\overline{V}}(L)}
{\exp\left\{- \frac{\frac{\sigma_s^2}{\sigma_w^2}\|\mathbf{y}\|^2+\|\mathbf{y}+\mathbf{x}\|^2}{2\frac{\sigma_s^4}{\sigma_s^2+\sigma_w^2}}\right\}} d\mathbf{y}
\nonumber\\
&= \frac{\frac{1}{({2\pi}\alpha\sigma_s\sigma)^{n_L}}}{f_{\sigma_s,\mathbf{c}}(L)} \times \nonumber\\
&\quad \quad  \int_{\mathcal{\overline{V}}(L)}
{\exp\left\{- \frac{\|\mathbf{y}\|^2}{2\tilde{\sigma}_w^2}\right\} \sum_{\mathbf{x}\in L-\mathbf{c}} \exp\left\{- \frac{\|\mathbf{y}+\mathbf{x}\|^2}{2\frac{\sigma_s^4}{\sigma_s^2+\sigma_w^2}}\right\}} d\mathbf{y}
\label{eq:Pe-average}
\end{align}
where we recall the definition $\tilde{\sigma}_w\triangleq \frac{\sigma_s\sigma}{\sqrt{\sigma_s^2+\sigma_w^2}}$ in the last step.

Now the key observation is that, by Lemma \ref{lem:2}, the infinite sum over $L-\mathbf{c}$ within the above integral is almost a constant for any $\mathbf{y}$ and any $\mathbf{c}$, as described in (\ref{eqn-dbl-1}) shown at the top of next page.

Substituting (\ref{eqn-dbl-1}) back into (\ref{eq:Pe-average}), and noting that $f_{\sigma_s,\mathbf{c}}(L) \in [1-\epsilon_{L}\left(\sigma_s\right),1+\epsilon_{L}\left(\sigma_s\right)] \frac{1}{V(L)}$, we derive the expression of $P_e$ as shown in (\ref{eqn-dbl-2}) at the top of next page, where (a) holds under the conditions $\epsilon_{L}\left(\frac{\sigma_s^2}{\sqrt{\sigma_s^2+\sigma_w^2}}\right) \rightarrow 0$ and $\epsilon_{L}\left(\sigma_s\right)\rightarrow0$.

But (\ref{eqn-dbl-2}) is just the error probability of standard lattice decoding for noise variance $\tilde{\sigma}_w^2$, previously studied by Poltyrev \cite{Poltyrev94}.

\begin{figure*}[!t]
\normalsize
\setcounter{mytempeqncnt}{\value{equation}}

\setcounter{equation}{34}
\begin{equation}
\label{eqn-dbl-1}
\sum_{\mathbf{x}\in L-\mathbf{c}} \exp\left\{- \frac{\|\mathbf{y}+\mathbf{x}\|^2}{2\frac{\sigma_s^4}{\sigma_s^2+\sigma_w^2}}\right\} \in \left[1 - \epsilon_{L}\left(\frac{\sigma_s^2}{\sqrt{\sigma_s^2+\sigma_w^2}}\right), 1 + \epsilon_{L}\left(\frac{\sigma_s^2}{\sqrt{\sigma_s^2+\sigma_w^2}}\right)\right] \left(\sqrt{2\pi}\frac{\sigma_s^2}{\sqrt{\sigma_s^2+\sigma_w^2}}\right)^{n_L} \frac{1}{V(L)}.
\end{equation}

\begin{align}\label{eqn-dbl-2}
P_e &\in \left[\frac{1 - \epsilon_{L}\left(\frac{\sigma_s^2}{\sqrt{\sigma_s^2+\sigma_w^2}}\right)}{1+ \epsilon_{L}\left(\sigma_s\right)},\frac{1 + \epsilon_{L}\left(\frac{\sigma_s^2}{\sqrt{\sigma_s^2+\sigma_w^2}}\right)}{1- \epsilon_{L}\left(\sigma_s\right)}\right] {\frac{1}{({2\pi}\alpha\sigma_s\sigma)^{n_L}}} \left(\sqrt{2\pi}\frac{\sigma_s^2}{\sqrt{\sigma_s^2+\sigma_w^2}}\right)^{n_L} \int_{\mathcal{\overline{V}}(L)}
\exp\left\{- \frac{\|\mathbf{y}\|^2}{2\tilde{\sigma}_w^2}\right\} d\mathbf{y}
\nonumber\\
&= \left[\frac{1 - \epsilon_{L}\left(\frac{\sigma_s^2}{\sqrt{\sigma_s^2+\sigma_w^2}}\right)}{1+ \epsilon_{L}\left(\sigma_s\right)},\frac{1 + \epsilon_{L}\left(\frac{\sigma_s^2}{\sqrt{\sigma_s^2+\sigma_w^2}}\right)}{1- \epsilon_{L}\left(\sigma_s\right)}\right]
\frac{1}{\left(\sqrt{2\pi}\tilde{\sigma}_w\right)^{n_L}} \int_{\mathcal{\overline{V}}(L)}
\exp\left\{- \frac{\|\mathbf{y}\|^2}{2\tilde{\sigma}_w^2}\right\} d\mathbf{y} \nonumber\\
&\stackrel{(a)}{\to} \frac{1}{(\sqrt{2\pi}\tilde{\sigma}_w)^{n_L}} \int_{\mathcal{\overline{V}}(L)}
\exp\left\{- \frac{\|\mathbf{y}\|^2}{2\tilde{\sigma}_w^2}\right\} d\mathbf{y}.
\end{align}
\setcounter{equation}{\value{mytempeqncnt}}
\hrulefill
\vspace*{4pt}
\end{figure*}

\footnotesize
\bibliographystyle{IEEEtran}
\bibliography{IEEEabrv,lingbib}

%

\begin{IEEEbiographynophoto}{Cong Ling}
received the B.S. and M.S. degrees in electrical engineering from
the Nanjing Institute of Communications Engineering, Nanjing, China,
in 1995 and 1997, respectively, and the Ph.D. degree in electrical
engineering from the Nanyang Technological University, Singapore, in
2005.

He is currently a Senior Lecturer in the Electrical and Electronic
Engineering Department at Imperial College London. His research
interests are coding, signal processing, and security, especially lattices.
Before joining Imperial College, he had been on the
faculties of Nanjing Institute of Communications Engineering and
King's College.

Dr. Ling is an Associate Editor of IEEE Transactions on Communications.
He has also served as an Associate Editor of IEEE Transactions on Vehicular Technology.
\end{IEEEbiographynophoto}

\begin{IEEEbiographynophoto}
{Jean-Claude Belfiore} (M’91) received the ``Dipl\^{o}me
d'ing\'{e}nieur" (Eng. degree) from Ecole Sup\'{e}rieure
d'Electricit\'{e} (Supelec) in 1985, the ``Doctorat" (PhD) from ENST
in 1989 and the ``Habilitation \`{a} diriger des Recherches" (HdR)
from Universit\'{e} Pierre et Marie Curie (UPMC) in 2001. In 1989,
he was enrolled at the ``Ecole Nationale Sup\'{e}rieure des
T\'{e}l\'{e}communications", ENST, also called ``T\'{e}l\'{e}com
ParisTech", where he is presently full Professor in the
Communications and Electronics department. He is carrying out
research at the Laboratoire de Traitement et Communication de
l'Information, LTCI , joint research laboratories between ENST and
the ``Centre National de la Recherche Scientifique" (CNRS), UMR
5141, where he is in charge of research activities in the areas of
digital communications, information theory and coding. Jean-Claude
Belfiore has made pioneering contributions on modulation and coding
for wireless systems (especially space-time coding) by using tools
of number theory. He is also, with Ghaya Rekaya and Emanuele
Viterbo, one of the co-inventors of the celebrated Golden Code. He
is now working on wireless network coding, coding for physical
security and coding for interference channels. He is author or
co-author of more than 200 technical papers and communications and
he has served as advisor for more than 30 Ph.D. students. Prof.
Belfiore has been the recipient of the 2007 Blondel Medal. He is an
Associate Editor of the IEEE Transactions on Information Theory for
Coding Theory.
\end{IEEEbiographynophoto}




\end{document}